# Developing an Efficient DMCIS with Next-Generation Wireless Networks

Al-Sakib Khan Pathan and Choong Seon Hong, *Member, IEEE*

*Abstract*—The impact of extreme events across the globe is extraordinary which continues to handicap the advancement of the struggling developing societies and threatens most of the industrialized countries in the globe. Various fields of Information and Communication Technology have widely been used for efficient disaster management; but only to a limited extent though, there is a tremendous potential for increasing efficiency and effectiveness in coping with disasters with the utilization of emerging wireless network technologies. Early warning, response to the particular situation and proper recovery are among the main focuses of an efficient disaster management system today. Considering these aspects, in this paper we propose a framework for developing an efficient Disaster Management Communications and Information System (DMCIS) which is basically benefited by the exploitation of the emerging wireless network technologies combined with other networking and data processing technologies.

## I. Introduction

THE increasing complexity of societies and growing specialization in hazard management clearly demonstrates that no authority or discipline could identify and address all of the significant consequences of hazards. This applies whether hazards be natural (earthquakes, extreme weather events, etc.), human-induced (such as nuclear or hazardous chemical accidents), or the interaction of both. The impact of hazards cuts across economic, social and political divisions in society so that the adequacy of the cumulative response is greatly influenced by the degree to which proactive as well as reactive actions can be effectively integrated and optimized. Successful integration of hazard reduction efforts, however, depends on the ability of organizations and individuals involved in all phases of the disaster management process (prevention, preparedness, response, and recovery/ reconstruction) to work together to develop and implement solutions to commonly recognized problems. In this regard, key factor in effective mitigation is the information and communication infrastructure that contributes to building knowledge about hazards and the interpretive processes, which in turn contributes to the formulation of options for collective action. The basic building block of the communications infrastructure is formed by various types of networks. The staggering growth of the wireless networks, plummeting costs of various types of telecommunications devices, and emerging next-generation wireless technologies add new dimension as well as show great promise for efficient and faster deployable information and communications infrastructure.

The intent of this paper is to explore the potential of the next-generation wireless networks to develop an efficient Disaster Management Communications and Information System (DMCIS) which basically relies on the utilization of next generation wireless networks. In addition to this, the key issues for developing an efficient DMCIS are also discussed.

This paper is organized as follows: Following the section I, Section II gives an overview of Disaster Management Communications and Information System (DMCIS), Section III introduces the emerging wireless network technologies in brief and presents our proposed framework in detail, Section IV gives an analysis of the proposed system, Section V mentions some of the related works and Section VI concludes the paper mentioning the future works to be done.

## II. Disaster Management Communications and Information System (DMCIS) – An Overview

It takes immense supremacy and courage to confront the situations, when man-made or natural disasters, such as earthquakes, floods, plane crashes, high-rise building collapses, major nuclear facility malfunctions etc. occur. In order to cope with such disasters in a fast and highly coordinated manner, the optimal provision of information concerning the situation is an essential pre-requisite. Police, fire departments, public health, civil defense and other organizations have to react not only efficiently and individually, but also in a coordinated manner [1]. For establishing a controlled system, various types of information need to be stored in and inter-communicated among the various hierarchy levels. Thus, the requirements for an integrated communications and information system for

Manuscript received April 26, 2006. This work was supported in part by the MIC and ITRC projects. Dr. C. S. Hong is the corresponding author.
Al-Sakib Khan Pathan is a graduate student and research assistant in the Networking Lab, Department of Computer Engineering, Kyung Hee University, South Korea (phone: +82 31 201-2987; fax: +82 31 204-9082; e-mail: spathan@networking.khu.ac.kr).
Dr. Choong Seon Hong is a professor in the Department of Computer Engineering, Kyung Hee University, South Korea (phone: +82 31 201 2532; fax: +82 31 204-9082; e-mail: cshong@khu.ac.kr).

managing disasters becomes an essential need for providing efficient, reliable and secure exchange and processing of relevant information.

Over the course of the past decade, tremendous changes to the global communication infrastructure have taken place, including the popular uptake of the Internet, the rapid growth and reduction of costs of mobile telecommunications, and the implementation of advanced space-based remote sensing and satellite communication systems. With these technologies recent advancements in the areas of the emerging wireless ad hoc and wireless sensor network technologies promise to transform the field of disaster management with an ambitious goal to enhance planning and reduce loss of life and property through improved communications.

In effect, two major developments have taken place within the last decade: a conceptual shift in disaster management towards more holistic and long-term risk reduction strategies, and a communication revolution that has increased dramatically both the accessibility of information and the functionality of communication technology for disaster management. While these shifts hold great promise for significantly reducing the impact of disasters, many issues remain to be addressed or resolved [2], [3], [4]. These include risk management and sustainable development, emergency telecommunications policy and appropriate technology transfer.

Two major categories, different but closely dependant on each other, involved in a Disaster Management Communications and Information System are:

--*Pre-disaster activities:* analysis and research (to improve the existing knowledge base), risk assessment, prevention, mitigation and preparedness

--*Post-disaster activities:* response, recovery, rehabilitation, and reconstruction.

Accordingly, there are two categories of disaster-related data:

--Pre-disaster baseline data about the location and risks or warning data

--Post-disaster real-time data about the impact of hazard and the resources available to combat it

Decision making of disaster management, having done proper risk analysis and discussion upon appropriate counter measures, can be greatly enhanced by the cross-sectional integration of information. For example, to understand the full short and long term implications of floods and to plan accordingly requires the analysis of combined data on meteorology, topography, soil characteristics, vegetation, hydrology, settlements, infrastructure, transportation, population, socio-economics and material resources. These information come from many different sources and it is often difficult in most countries to bring them all together.

The components of a DMCIS involves a number of databases that store various sorts of data and information about vulnerability assessment, demographic data, available resources, disaster impact factor of various types of disasters etc. The usage of DMCIS could be in three contexts

--Preparedness planning

--Mitigation

--Response & recovery

For all of these tasks, the primary task is to provide valid, accurate and timely data from the disaster affected areas to the decision making centers which in turn would take the measures for mitigation and recovery of the loss caused by disasters. Efficient networks could play the major role for this data hunting and co-ordination. In fact, next-generation wireless networks could effectively be used for this purpose.

III. AN EFFICIENT DMCIS AIDED WITH EMERGING WIRELESS NETWORKS

*A. Wireless Sensor and Ad Hoc Networks*

Wireless sensor network and wireless ad hoc networks are two emerging technologies that show great promise for various futuristic public applications. In this sub-section, we mention the major characteristics of these two networks in brief.

Wireless sensor network [5], [6] is a combination of hundreds and thousands of small sensing devices or sensors, also known as wireless integrated network sensors (WINS) [7]. An integration of sensing circuitry, processing power, memory and wireless transceiver makes a smart wireless sensor. These tiny devices are considered to contribute significantly to the field of networking and expected to be used in abundance for various practical purposes in future as they are suitable for sensing the change of many significant parameters like heat, pressure, light, sound, soil makeup, movement of objects etc. The sensors could use acoustic, seismic, infrared, thermal or visual mechanism for detecting the incidents. Hence a wireless sensor network consisting of these devices could effectively be used in the hazardous environments for moving object detection, environmental monitoring, surveillance etc. especially for collecting warning data from natural or in some cases from human-induced disasters.

Wireless ad hoc networks, on the other hand are self-organizing, dynamic topology networks formed by a collection of mobile nodes through radio links [8]. Minimal configuration, absence of infrastructure and quick deployment make them convenient for emergency situations. Major features of wireless ad hoc networks are the wireless connectivity, mobile nodes in the network, ease of deployment, speed of deployment and anytime, anywhere deployment e.g. ad hoc nature.

*B. Details of our Framework*

To achieve the goal of timely, accurate and reliable data collection about disaster hotspots and emergency situations we use both wireless ad hoc and wireless sensor networks in our framework. The framework is basically divided into four distinct levels which interact with each other to increase the efficiency of the Disaster Management Communications and Information System. In this section, we present the details of our proposed framework.

*Level One – Deployment of Sensors and SDCCs.* Wireless sensor networks underpin the level one of the framework. The primary task of level one is to collect raw data (e.g. sensor data) using the wireless sensor networks. Sensors are deployed in the crucial parts of the disaster prone areas like river banks, seashore, hilly areas and other areas of interest. The sensors monitor the change of certain parameters and if significant changes are detected (e.g., level of water in the rivers which could help for flood warning, earthquakes, tsunami, cyclones etc.) they send them to the sink or Sensor Data Collection Center (SDCC). Each of the disaster-prone areas has at least one SDCC located nearby, which is equipped with computers for storing acquired data. These SDCCs are also equipped with wireless transceivers. Here, we assume that, the wireless sensors remain relatively fixed once they are deployed. The sensors in the network could be of different types (acoustic, magnetic, seismic etc.). A clustering approach like [9] could be used for the formation of clusters of sensors. Though the major task of SDCC is to store data, SDCCs could incorporate some local data processing mechanisms. For such processing, there is a threshold value $\tau$ representing the number of sensors that should send their sensor data to the SDCC. Let, each sensor is assigned an id $i$ and is represented by $s_i$ where $i=1,2,3,....N$. Hence, for a particular area assigned to one SDCC,

$$\tau \leq \sum_{i=1}^{N} s_i \quad where, \sum_{i=1}^{N} s_i \gg 1 \quad (1)$$

In addition to the data collected by the wireless sensor networks, other necessary data like demographic data, health care information, available facilities and resources in the area etc. could be manually inserted into the SDCC. Figure 1 shows level one data acquisition phase.

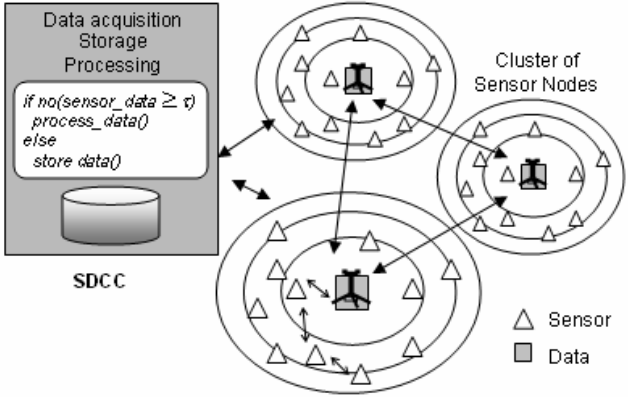

Fig. 1. Data Acquisition using Wireless Sensor Networks

*Level Two – Wireless Ad Hoc Networks for Data Transmission to DPC.* In level two, Mobile Access Points (MAPs) play the major role. A MAP is a vehicle mounted wireless access point which uses low-cost Wi-Fi (Wireless Fidelity) technology [10], [11]. When a MAP comes near to an SDCC, a wireless ad hoc network is automatically formed and all the raw data or partially processed data are downloaded into the MAP. The 802.11b Wi-Fi technology operates in the 2.4 GHz range offering data speeds up to 11 megabits per second [12]. There are two other specifications that offer up to five times the raw data rate, or 54 Mbps. One is 802.11g which operates on the same 2.4 GHz frequency band as 802.11b. The other alternative 802.11a, occupies frequencies in the 5 GHz band. If offers less range of coverage than either 802.11b and 802.11g but offers up to 12 non-overlapping channels, compared to three for 802.11b or 802.11g, so it can handle more traffic than it's 2.4 GHz counterpart [13]. Any of these specifications is chosen for a particular area for transmitting stored data from an SDCC to the MAPs. These data are then taken by the MAPs and delivered to the DPCs (Data Processing Center) located at nearby areas; which are considered to be relatively safer from the disaster hotspots.

Several $MAP_j$ where $j=1,2,3,...J$ ($J$ is the maximum limit of the ids of the MAPs) are associate with each pair of SDCC and Data Processing Center (DPC).

Here, $\sum_{j=1}^{J} MAP_j \geq \sum_{r=1}^{R} SDCC_r$

and, $\sum_{j=1}^{J} MAP_j \geq \sum_{t=1}^{T} DPC_t$

but not necessarily, $\sum_{r=1}^{R} SDCC_r = \sum_{t=1}^{T} DPC_t \quad (2)$

Here, T and R are the maximum numbers of DPC and SDCC ids for a particular region. SDCC and DPC pairs could be crossly inter-connected or overlap for reliability of the acquired data. The MAPs move around the SDCC(s) and collect data from them.

So, a Wi-Fi enabled MAP operates in two ways:
1) Forms wireless ad hoc network when comes close to the SDCC and collects data from the SDCC using Wi-Fi radio transceivers.
2) Again, forms wireless ad hoc network when comes close to the DPCs and delivers raw data (or partially processed data) to the DPCs using Wi-Fi radio transceivers.

The major task of the MAPs is to ensure quick acquisition and delivery of raw or partially processed data about possible disaster and to bridge the gap between the areas under threat and the safer areas, from where the warning, preparedness and recovery instructions would be issued.

Now, if the SDCC and DPC represent the same center e.g.
if, $SDCC_r = DPC_t$ for some $r$ and $t$
or, if $\| SDCC_r, DPC_t \| < \delta$

where $\delta$ is the minimum threshold distance needed to be maintained between a DPC-SDCC pair and $\|x, y\|$ denote the distance between two locations $x$ and $y$, then Mobile Access Points (MAP) are not required for data collection and delivery as, the concerned SDCC and DPC can directly communicate using their wireless transceivers or within it. Also formation of ad hoc networks is not necessary for the same reason.

*Level Three – Processing Acquired Data in Data Processing Centers.* As DPCs are capable of wireless communications all the incoming data from the MAPs are stored and processed in the DPCs. For data integrity and authenticity, all the DPCs are networked among themselves using wireless or wired connections or combination of both. A DPC also contain detailed past records or history of disasters for the areas which are associated with it. For certain types of disasters the past records are very crucial to judge or estimate the risk of occurrence of that type of disaster again. For example, issuing a flood warning not only needs current data about the water level, flow path of the river, characteristics of the river etc. but also needs to compare all of these information with the records of past few years for that particular region.

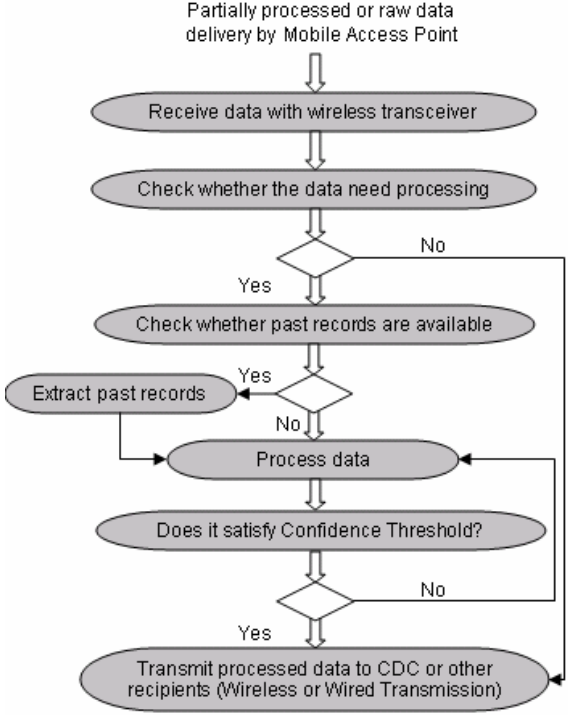

Fig. 2. Operation Process for DPC

This data processing step could be minimal in case of some disasters like, earthquake, building collapse as these sorts of disasters could happen without proper notice and hence, response, recovery and relief become most important tasks. The operation process of a DPC is shown in Figure 2. In this level, the related data from other DPCs could also be collected using wired or wireless transmission. After processing the data, confidence threshold is checked. Error detection is defined by the predetermined threshold and if necessary sent back to processing mechanism for further processing. Once processed data is ready, they are transmitted to the CDC (Central Data Center) for the next level. Wireless or wired transmission could be used for this data transfer.

*Level Four – Data Distribution and Response.* In this level, the processed data from the DPCs are gathered in the CDC. Of course,

$$\sum_{a=1}^{A} \sum_{t=1}^{T} DPC_t \gg c \qquad (3)$$

Where $c$ is the number of CDC(s) and $a$ is the particular area id.

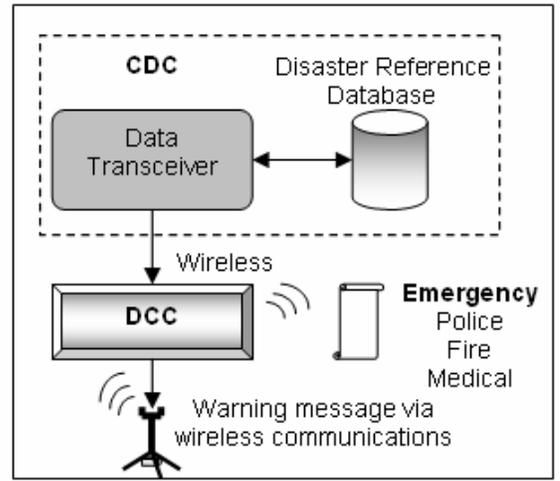

Fig. 3. CDC Module and DCC using Wireless Networks

Now, the task of CDC is to check the similarities of the information with past records of already occurred disasters for that particular area. A reference database of past disasters helps in this case. Depending upon finding the probability of occurring similar events, CDC requests DCC (Decision and Command Center) for taking responsive actions which in turn could call the emergency departments like police, fire, medical etc. Figure 3 represents a CDC module. In this way, timely, processed and accurate data from the target areas could be supplied and necessary preventive or reactive actions could be taken. To warn people about the possibilities of imminent disaster(s), the DCC sends information to the local mobile phone service providers to disseminate the warning via SMS (Short Message Service) to their mobile subscribers. This sort of warning could be very helpful for facing the disasters like cyclones or tsunamis. In addition to this, the DCC could also use Internet messaging or other web services.

Depending upon the gravity of the data, for example, emergency situations like, tornados, flash floods, earthquakes, landslides, building collapse etc., the MAPs or the DPCs could use the wireless communications to directly call the emergency or other services bypassing the CDC and DCC.

## IV. ANALYSIS OF OUR FRAMEWORK

In this section, we analyze our framework for determining the efficiency of the DMCIS. The initial task of raw data collection is done by wireless sensor networks and at least $\tau$ number of sensor inputs is necessary. As mentioned in equation (1), in level one, $\tau \leq \sum_{i=1}^{N} s_i \quad where, \sum_{i=1}^{N} s_i \gg 1$. Depending upon the type of disaster to be dealt with, the value of $\tau$ could be set very close to the value of $\sum_{i=1}^{N} s_i$ or equal. For flood or tsunami warning, it could be required that each and every sensor in the total wireless sensor network in that region must contribute for better data analysis. This is done to make sure that wrong reports do not lead to issuing a wrong warning by the CDC. As the sensors are deployed along the river banks and seashores, some of them might send wrong readings about the water levels. In fact, only a rise of water-level of a particular part of the river or sea may not cause flood or tsunami. For example; water could be agitated by the movements of the boats/ships and thus could lead the threshold number ($\tau$) of sensors to report wrongly. In such cases, $\tau$ should be set the value as close as possible to the value of N or exactly N. However, if we consider the failure or damage of some of the sensor nodes, it becomes impractical to set the value of $\tau$ exactly equal to $\sum_{i=1}^{N} S_i$. Also, to prevent the generation of false data in level one, detailed analysis should be done before deployment of the wireless sensors in their respective positions.

TABLE I
PROPOSED FRAMEWORK AT A GLANCE

| Levels | Type of Comm. | Major Contributors | Disaster Data | Necessary Conditions |
|---|---|---|---|---|
| Level One | Wireless and Wired | Wireless Sensor Networks | Raw or Processed Data (Depends upon the situation at hand) | $\tau \leq \sum_{i=1}^{N} S_i$ where, $\sum_{i=1}^{N} S_i \gg 1$ |
| Level Two | Wireless | Ad Hoc Networks and Mobile Access Points | Data downloaded from SDCCs | $\sum_{j=1}^{J} MAP_j \geq \sum_{r=1}^{R} SDCC_r$ and, $\sum_{j=1}^{J} MAP_j \geq \sum_{t=1}^{T} DPC_t$ but not necessarily, $\sum_{r=1}^{R} SDCC_r = \sum_{t=1}^{T} DPC_t$ |
| Level Three | Wireless and Wired | Data Processing Centers | Inputs from MAP, Outputs are fully processed data | |
| Level Four | Wireless and Wired | Central Data Center and Decision and Command Center | Processed data for reference and Warning message | $\sum_{a=1}^{A} \sum_{t=1}^{T} DPC_t \gg c$ |

In level two Mobile Access Point carriers are used because; in many cases it is difficult to set up wired networks for doing the same task. The using of MAPs could definitely increase the cost-efficiency for raw data transmission to DPCs. The DPCs are set up in the areas considered relatively safer from the areas where SDCCs are located. The formation of wireless ad hoc networks between SDCC and MAP or MAP and DPC could efficiently transfer data. Once the data are received by the DPCs, fully processed, confidence threshold is checked and sent to the CDC, CDC does not have to think about the fidelity of the data. It can quickly check the disaster reference database, store the data as future reference and send request to the DCC which in turn sends the warning messages to all the concerned units. Table I represents various facets of our framework at a glance. The use of the existing mobile phone networks for delivering SMS is a cost-effective solution to provide warning messages to the public, prior happening most of the natural disaster.

Also the condition, $\sum_{a=1}^{A} \sum_{t=1}^{T} DPC_t \gg c$ Indicates that as the total number of DPCs is huge compared to that of CDC, data processing could be done in several DPCs in parallel when the amount of data is huge. In fact, disasters like tsunami or earthquake in the hilly areas require a lot of data to be processed before an appropriate action could be taken. As a whole we believe that, our framework promises fast delivery and processing of data as well as could warn people about possible disastrous situations in advance using wireless communications.

## V. RELATED WORKS

[14] discusses some general issues on disaster management and presents a disaster area architecture which was developed to crystallize and capture information requirements for emergency support functions managers in the field. Lee et. al. [15] described a template-driven design methodology for disaster management information system which aims at archiving past-disaster relief operations. This template-based methodology could be useful for generating the disaster reference database. Kuwata et. al. [16] propose a work flow model for a disaster response system which consists of mainly four steps: data acquisition, data analysis, decision support and command and control. In [17] the author reviews emergency telecommunications and explores the role of information and communication technologies for disaster mitigation and humanitarian relief. In [18] the authors present a framework for data collection using sensor networks in disaster situations. Kamegawa et. al. [19] developed an

algorithm termed ADES which can share victims' information with all shelters and the purpose of their work is to form a wireless network system for particularly rescue activities in the distressed areas.

Our work differs from all of these works as we propose a detailed framework which incorporates various emerging wireless networks and database systems to work together for disaster prevention, mitigation and damage control. Also it is possible to implement the framework for a specific type of disaster.

## VI. CONCLUSIONS

Communications and Information Technologies, skills, and media are essential to link scientists, disaster mitigation officials, and the public; to educate the public about disaster preparedness; to track approaching hazards; to alert authorities; to warn the people most likely to be affected; to assess damage; to collect information, supplies, and other resources; to coordinate rescue and relief activities; to account for missing people; and to motivate public, political or institutional responses.

In this paper we proposed a detailed framework of an efficient Disaster Management Communication and Information System which takes the advantage of the next-generation wireless networks. While the networks would help for better, quick and reliable data delivery from the disaster hotspots, other associated technologies like disaster prediction or forecasting, databases, web services, intelligent systems, image processing etc. should work collaboratively for tackling disasters successfully. Moreover, acquiring secured data at every step is very crucial. The base of our framework is the wireless sensor networks (WSN) and security in WSNs is still a hot research issue. We are currently working on secured transmission of data at each level and between two distinct levels, starting from the level one of our framework.